\documentclass[
    ,final            
  ]
  {aipproc}
\usepackage{natbib}
\layoutstyle{8x11double}

\begin{document}

\title{Implications of kHz QPOs for the \\Spin Frequencies and Magnetic Fields of Neutron Stars:\\ New Results from Circinus X-1}


\author{Stratos Boutloukos }{
  address={Center for Theoretical Astrophysics and Department of Physics, University of Illinois, 1110 W. Green St., Urbana, IL 61801, USA},
  ,email={stratos@uiuc.edu}
}

\author{Frederick K. Lamb }{
  address={Center for Theoretical Astrophysics and Department of Physics, University of Illinois, 1110 W. Green St., Urbana, IL 61801, USA},
  altaddress={Also, Department of Astronomy},
  email={fkl@uiuc.edu}
}

\begin{abstract}
Detection of paired kilohertz quasi-periodic oscillations (kHz QPOs) in the X-ray emission of a compact object is compelling evidence that the object is an accreting neutron star. In many neutron stars, the stellar spin rate is equal or roughly equal to $\Delta\nu$, the frequency separation of the QPO pair, or to $2\Delta\nu$. Hence, if the mechanism that produces the kilohertz QPOs is similar in all stars, measurement of $\Delta\nu$ can provide an estimate of the star's spin rate. The involvement of the stellar spin in producing $\Delta\nu$ indicates that the magnetic fields of these stars are dynamically important.

We focus here on the implications of the paired kHz QPOs recently discovered in the low-mass X-ray binary (LMXB) system Cir X-1 \citep{cirx1}. The kHz QPOs discovered in Cir X-1 are generally similar to those seen in other stars, establishing that the compact object in the Cir X-1 system is a neutron star. However, the frequency $\nu_u$ of its upper kHz QPO is up to a factor of three smaller than is typical, and $\Delta\nu$ varies by about a factor 2 (167 Hz, the largest variation so far observed). Periodic oscillations have not yet been detected from Cir X-1, so its spin rate has not yet been measured directly. The low values of $\nu_u$ and the large variation of $\Delta\nu$ challenge current models of the generation of kHz QPOs. Improving our understanding of Cir~X-1 will improve our knowledge of the spin rates and magnetic fields of all neutron stars.
\end{abstract}

\maketitle

\section{Current understanding}\label{sec:1}

  Power density spectra of the X-ray brightness of LMXBs reveal various types of QPOs. Some two dozen of the 187 known LMXBs \citep{lmxbs} have been observed to produce simultaneously two QPOs with frequencies $\sim100-1300$ Hz that increase and decrease together by hundreds of Hz \citep{pbk}. Some of the key observational features of these kHz QPOs and their connection to our current theoretical understanding can be summarized as follows \citep{calgary-review}:
\begin{enumerate}
\item The frequency of one of the two kHz QPOs probably reflects the azimuthal frequency of gas in a streamline at a particular radius in the inner disk. The frequencies of the kHz QPOs are similar to orbital frequencies near a neutron star and vary by hundreds of Hertz on time scales as short as minutes. Such large, rapid variations are possible if the frequency is related to the orbital frequency at a radius that varies. This can be used to constrain the equation of state of neutron stars \citep{mlp98}.
\item The star's spin appears to be involved in producing the frequency separation $\Delta\nu$ of the two kilohertz QPOs in a pair. This is evident in the millisecond pulsars (MSPs) XTE~J1807$-$294, where $\Delta\nu\approx \nu_s$ \citep{linares-05}, and SAX~J1808.4$-$3658, where $\Delta\nu\approx \nu_s/2$ \citep{Wijnands-03}. It is strongly indicated in the other kilohertz QPO sources, because in all cases where both $\Delta\nu$ and $\nu_s$ have been measured, the largest value of $\Delta\nu$ is consistent or approximately consistent with either $\nu_s$ or $\nu_s/2$ \citep{vdKlis06}. Thus, if paired kHz QPOs are detected but accretion- and nuclear-powered X-ray oscillations are not, a rough estimate of the star's spin rate can be made using $\Delta\nu$.
\item The spin rate may be communicated directly by the star's magnetic field or indirectly, e.g., via an anisotropic radiation pattern that rotates with the star. In eihter case the star's magnetic field must be dynamically important.
\item In order to explain the oscillation amplitudes observed, the X-rays must be coming from the surface of the neutron star, because the energy available at larger radii is far too small.
A viable model should explain (a) what determines the frequency (radius) at which the principal QPO is generated; (b) why the width of a kHz QPO peak (range of orbital radii) is so small; (c) how the frequency of the principal QPO changes by factors $\sim2-3$; and (d) how the spin frequency of the neutron star is involved. So far, the only model that may explain these key features of the kHz QPOs and their amplitudes, energy-dependence, and dependence on the mass accretion rate is the sonic-point spin-resonance model \citep{spsr2}.
\end{enumerate}  

\begin{figure}[t]
  \includegraphics[height=.25\textheight]{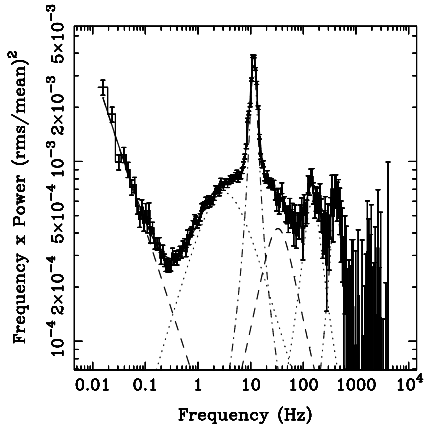}
  \caption{Spectral energy density of variations in the X-ray brightness of Cir X-1, from an archived RXTE observation. Right: 136 and 404 Hz ``kHz'' QPOs; middle: low-frequency QPO; left: very low frequency noise \citep{cirx1}.}\label{twins}
\end{figure}

\section{Discovery of paired kHz QPOs in Cir X-1}

\begin{figure}[b]
  \rotatebox{-90}{\includegraphics[height=.35\textheight]{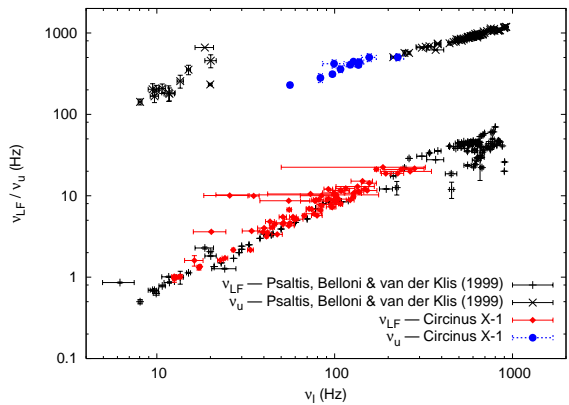}}
  \caption{Frequencies of the upper kHz QPO (top) and the low-frequency QPO (bottom) of Cir X-1 and other NS LMXBs, plotted against the frequency of the lower kHz QPO. From \citep{pbk}.}\label{pbk}
\end{figure}

\citet{cirx1,cirx1-err} discovered a kHz QPO pair in 11 archived RXTE observations of Cir X-1 and a single kHz QPO in 69 other observations. Figure \ref{twins} shows results from one of these observations. The two kHz QPOs are visible at 136$\pm$5 Hz and 404$\pm$2 Hz. Their frequencies track one another and the frequency of the low-frequency QPO in the same way as in other NS LMXBs (see Fig.~\ref{pbk}). This, and the fact that the frequencies and frequency ratio of the kHz QPOs vary greatly from one observation to another demonstrates that the Cir X-1 system: 
\begin{enumerate}
\item[a)] contains a NS, resolving a debate that had continued for more than two decades 
\item[b)] has a dynamically important magnetic field, as inferred from the current understanding of how the kHz QPOs are generated, and
\item[c)] has a spin frequency of several hundred Hz, based on the frequency separation of the kHz QPO pair.
\end{enumerate}

Table \ref{spins} shows the inferred spin rates of EXO~0748$-$676 and the 23 known accreting millisecond pulsars. Nuclear- and accretion-powered oscillations have been detected in 17 and 10 of them, respectively; kHz QPOs have been detected in 11 MSPs.

  \begin{table}[!t]
\caption {Accretion- and nuclear-powered millisecond (spin period $P_{s}$ $<10$~ms) pulsars and EXO~0748$-$676.}\label{spins}
\begin{tabular}{@{\extracolsep{\fill}}lll}
\hline
\noalign{\kern 2pt}
$\nu_{\rm spin}$~(Hz)\tablenote{Spin frequency inferred from periodic or nearly periodic X-ray oscillations. A:~accretion-powered millisecond pulsar. N: nuclear-powered millisecond pulsar. K: kilohertz QPO source.}	&Object&Reference\cr
\noalign{\kern 2pt}
\hline
\noalign{\kern 2pt}
1122\ \ \ \,NK\qquad\qquad	& \hbox{XTE~J1739$-$285}	& \cite{1122hz}\cr
\, 619\ \ \ \,NK\quad\qquad                 &\hbox{4U~1608$-$52}    & \cite{hartman}\cr
\, 611\ \ \ \,N\qquad\qquad 	& \hbox{GS~1826$-$238}		& \cite{thompson-05}\cr
\, 601\ \ \ \,NK\quad\qquad                 &\hbox{SAX~J1750.8$-$2900}    & \cite{kaaret02}\cr
\, 598\ \ \ \,A \qquad\qquad		&\hbox{IGR J00291$+$5934}	& \cite{2004ATel..353....1M}\cr
\, 589\ \ \ \,N\qquad\qquad            &\hbox{X~1743$-$29}     & \cite{x1743}\cr
\, 581\ \ \ \,NK\quad\qquad         &\hbox{4U~1636$-$53}    & \cite{zhang96,Wijnands-97,S98b}\cr
\, 567\ \ \ \,N\qquad\qquad      &\hbox{MXB~1659$-$298}    & \cite{rudy01}\cr
\, 550\ \ \ \,ANK\quad\qquad                 &\hbox{Aql~X$-$1}         & \cite{casella,zhang98}\cr
\, 530\ \ \ \,N\qquad\qquad			&\hbox{A~1744$-$361}   & \cite{sudip06}\cr
\, 524\ \ \ \,NK\quad\qquad                 &\hbox{KS~1731$-$260}   & \cite{smith97}\cr
\, 442\ \ \ \,AN\qquad\qquad			&\hbox{SAX~J1748.9$-$2021}	& \cite{Gavriil-07,diego,kaaret03}\cr
\, 435\ \ \ \,A\qquad\qquad     &\hbox{XTE~J1751$-$305}    &\cite{Mark02}\cr
\, 401\ \ \ \,ANK\           &\hbox{SAX~J1808.4$-$3658\quad}    &\cite{rudy-michiel-nature,chakrabarty98}\cr
\, 377\ \ \ \,A \qquad\qquad 		& \hbox{HETE~J1900.1$-$2455}	& \cite{morgan05}\cr
\, 363\ \ \ \,NK\quad\qquad                     &\hbox{4U~1728$-$34}       &\cite{S96}\cr
\, 330\ \ \ \,NK\quad\qquad     &\hbox{4U~1702$-$429}      &\cite{markwardt99}\cr
\, 314\ \ \ \,AN\quad\qquad     &\hbox{XTE~J1814$-$338}    &\cite{markwardt-swank03}\cr
\, 294\ \ \ \,NK\quad\qquad		&\hbox{IGR J17191$-$2821}	&\cite{markwardt294,kleinwolt}\cr
\, 270\ \ \ \,N\qquad\qquad                         &\hbox{4U~1916$-$05}       &\cite{galloway01}\\
\, 191\ \ \ \,AK\quad\qquad        &\hbox{XTE~J1807.4$-$294}    &\cite{markwardt03a,W06}\cr
\, 185\ \ \ \,A\qquad\qquad     &\hbox{XTE~J0929$-$314}\ \ \ \    &\cite{Gal02}\cr
\, 182\ \ \ \,A\qquad\qquad		&\hbox{SWIFT~J1756.9$-$2508}\ \ \ \	&\cite{Krimm}\cr
\, \, 45\ \ \ \,N\quad\qquad     &\hbox{EXO~0748$-$676}\ \ \ \      &\cite{2004ApJ...614L.121V}\cr
\hline
\end{tabular}
\end{table}

\section{Unique Properties of the paired kHz QPOs seen in Cir X-1}

\begin{figure}[t]
    \includegraphics[height=.22\textheight]{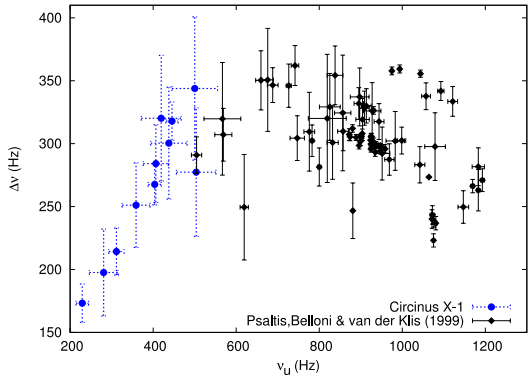}
      \caption{Frequency separation of kHz QPO pairs vs. frequency of the upper kHz QPO, for the same sources as in Fig.~\ref{pbk}.}\label{dv}
\end{figure}

The discovery of paired kHz QPOs in Cir X-1 has answered several important questions, but has also opened new ones:
\begin{itemize}
\item The frequencies of the paired kHz QPOs seen in Cir X-1 reach the lowest values so far observed in NS LMXBs: 55 Hz for the lower QPO and 250 Hz for the upper QPO. These observed frequencies extend the previously known frequency correlation (\citep{pbk}, Fig.~\ref{pbk}) over a range of a factor 4; single kHz QPOs were seen with frequencies as low as 10 Hz. According to the current theoretical understanding \ref{sec:1} (see above), 
this implies that the frequency of the principal kHz QPO is sometimes generated at a radius $\sim$50 km, for a 2 M$_\odot$ star.
\item The frequency separation $\Delta\nu$ of the kHz QPOs reaches the lowest value so far observed (170 Hz), and $\Delta\nu/\nu_u$ is $0.55-0.75$, larger than the values $\sim0.2-0.6$ observed in other kHz QPO systems.
\item The frequency separation varies by $\sim180$ Hz (more than a factor 2), more than the variation of the frequency separation among all other known systems (see Fig.~\ref{dv}).
\item The frequency separation increases systematically with increasing upper kHz QPO frequency, unlike in other kHz QPO sources.
\end{itemize}

The kHz QPOs discovered in Cir X-1 challenge our current theoretical understanding of how the kHz QPOs are generated. Understanding them will improve our understanding of how the kHz QPOs are generated, enhancing our ability to determine the masses, spin rates, and magnetic field of the neutron stars in LMXBs.

\begin{theacknowledgments}
These results are based on research supported by NSF Award numbers AST 00-98399 and AST 07-09015 and NASA Award number NAG 5-12030
\end{theacknowledgments}

\bibliographystyle{aipprocl}

\end{document}